\documentstyle[psfrag,graphicx,aps,twocolumn]{revtex}
\def\be{\begin{eqnarray}}
\def\ee{\end{eqnarray}}

\def\roughly#1{\mathrel{\raise.3ex\hbox{$#1$\kern-.75em%
\lower1ex\hbox{$\sim$}}}}

\newcommand{\eq}{\begin{equation}}
\newcommand{\eqx}{\end{equation}}
\newcommand{\eqn}{\begin{eqnarray}}
\newcommand{\eqnx}{\end{eqnarray}}

\begin{document}

\renewcommand{\thefootnote}{\arabic{footnote}}
\setcounter{footnote}{0}

\vskip 0.4cm

\title{\LARGE\bf Wealth Condensation in Pareto Macro-Economies}

\author{
Z. Burda$^{a,b}$, D. Johnston$^c$, 
J. Jurkiewicz$^a$, \\ M. Kami\'{n}ski$^d$,
M.A. Nowak$^a$, G. Papp$^e$ and I. Zahed$^f$}

\address{
$^a$ {\it M. Smoluchowski Institute of Physics,
Jagellonian University, Cracow, Poland} \\
$^b$ {\it Fakult\"at f\"ur Physik, Universit\"at Bielefeld
P.O.Box 100131, D-33501 Bielefeld, Germany} \\
$^c$ {\it Department of Mathematics, Heriot-Watt University, 
Edinburgh EH14 4AS, Scotland}\\
$^d$ {\it Fortis Bank Polska S.A.,  Postepu 15, 02-676 Warsaw,
Poland}\\
$^e$ {\it HAS Research Group for Theoretical Physics, 
E\"otv\"os University, Budapest, H-1518 Hungary}\\
$^f$ {\it Department of Physics and Astronomy,
State-University of New-York, Stony-Brook, NY 11794} 
}

\date{\today}
\maketitle
\begin{abstract}
We discuss a Pareto macro-economy (a) in a closed  system with fixed total
wealth and (b) in an open system  with average mean wealth and compare
our results to a similar analysis in a super-open system (c) with
unbounded wealth~\cite{bm}. Wealth condensation takes place in the
social phase for closed and open economies, while it occurs in the
liberal phase for super-open economies. In the first two cases, the 
condensation is related to a mechanism known from the balls-in-boxes
model, while in the last case to the non-integrable tails of the Pareto 
distribution. For a closed macro-economy in the social phase, we point 
to the emergence of a ``corruption'' phenomenon: a sizeable fraction of
the total wealth is always amassed by a single individual.

\end{abstract}



\vskip .2cm
{\bf 1.\,\,\,}
Power law distributions permeate a number of phenomena in statistical
physics and critical phenomena. They are an important manifestation of 
scale invariance as observed in fractals, self-organized criticality 
and percolating structures. Generically, they are the consequence
of the central limit theorem for scale free processes where a random
L\'{e}vy walk replaces Brownian motion~\cite{LEVY}. 

Power law distributions have also been suggested to describe social and
economic statistics. While the bulk of the income distribution in
most societies follows a log-normal distribution, about a century
ago Pareto suggested that the wealthy are outliers. The distribution
of large wealths follow a power law
\begin{equation}
p(w) \sim w^{-1 -\alpha} \quad {\rm for } \ w \gg w_0 \, .
\end{equation}
with $\alpha$ typically between 1-2. This distribution is referred
to as Pareto's distribution~\cite{p,m}. 
Power-like tails also govern the distribution of income and size
of firms, and the behavior of financial time series over intermediate
time horizons~\cite{b}. 
The scale free character of this distribution implies that the
chance of an already rich individual ($w \gg w_0$) to further increase 
his wealth by an additional factor $\lambda$ is $p(\lambda\,w)/p(w)\approx 
1/\lambda^{1+\alpha}$, independently of his current wealth and the
wealth of the less fortunate. For the rich part of the ensemble 
what matters is only the index $\alpha$ and, as we will argue,
the total wealth of the society.

A social engineer may attempt to use the value of $\alpha$ 
to control  the likelihood of large wealths in general, for 
instance by increasing the global character of trade through 
interest rates or decreasing it through taxation. The larger $\alpha$, the stronger  the 
suppression of large wealths. One can distinguish
between two separate regimes, the liberal economies with $\alpha \le 1$ and the
social ones with $\alpha > 1$. As we will see, the possibilities for
a condensation of wealth to occur are very different in both.

The total wealth $W$ of an economy, or alternatively the average wealth
per individual ${W}/{N}$, can become important as an upper bound
on the individual wealths $w$. For example, it is clear that there will
be no rich individuals in an uniformly poor society, even if the economy is
liberal ($\alpha \le 1$). Conversely, one can ask about what happens in
a rich society with a restrictive social economy ($\alpha > 1$). As we
will show, a Pareto macro-economy becomes unstable in this case and
favors a ``corrupt'' scenario where one individual amasses almost all the
available wealth.

To better understand the role of the macro-economic parameter ${W}/{N}$,
we now define the three advertised ensembles:
(a) a closed economy with a total wealth $W$ fixed;
(b) an open economy in equilibrium with external economies
where $W$ adjusts to the equilibrium mean; (c) a super-open
economy where $W$ can grow unrestricted.
From the point of view of rich individuals, the essential parameters of
the respective ensembles are the number $N$ of individuals
in the society which is kept fixed in all cases and:
(a) the Pareto index $\alpha$ and the average individual wealth
$w = {W}/{N}$ beyond a critical value  $w_*$ (see below);
(b) the Pareto index $\alpha$ and a stability parameter
$\mu$ (see below); (c) the Pareto index $\alpha$ only.

{\bf 2.\,\,\,} The authors of \cite{bm} recently 
proposed a simple theoretical model
of a dynamical process of wealth flows which in equilibrium becomes a
Pareto macro-economy for the ensemble (c). In brief, 
the model is given by a set of stochastic equations that describe the
flow of wealth in an ensemble of $N$ individuals. Specifically, the time
evolution of each individual's wealth $w_i(t)$, $i = 1, \dots, N$, is
assumed to be described by a linear differential equation:
\begin{equation}
\frac{d w_i (t)}{d t} = \eta_i(t) w_i (t) 
+ \sum_{j (\ne i)}^{N} J_{ij}  w_j (t) 
- \sum_{j (\ne i)}^{N} J_{ji}  w_i (t) \, .
\label{dyn}
\end{equation}
Here, trading between individuals is encoded in the buy/sell flow channels
$J_{ij}$ that describe an internal macro-economical network. In
addition,  each individual is subjected to an
economical background which is given by a multiplicative random source
$\eta_i(t)$, representing the spontaneous increase or decrease of wealth
related to investments, gains and losses on the market, {\em etc}. By
construction, the equations are invariant under change in
monetary unit, $w_i \rightarrow \lambda w_i$.

In general, both $\eta_i(t)$ and $J_{ij}$ can be very complicated functions.
Following \cite{bm}, here we will discuss only the simplest case, where we assume that the
$\eta_i(t)$ are just uncorrelated random variables with a Gaussian
distribution, and that all interactions between individuals are the same,
$J_{ij} = {J}/{N}$ for all $i \ne j$ (mean-field). As a result, 
the corresponding equilibrium probability distribution has the following large-$w$ asymptotics:
\begin{equation}
p(w) \ \sim \ w^{- 1 - \alpha}
\label{pls}
\end{equation}
where $\alpha = 1 + {J}/{\sigma^2} > 1$ and $\sigma^2$ is the variance
of the Gaussian distribution of $\eta(t)$. The normalization factor, which
we left out of (\ref{pls}) for simplicity, depends on $\alpha$ only.
For large values of $w$, this solution gives a power law with $\alpha > 1$,
{\em i.e.} we are in the generic situation of a social Pareto macro-economy.
However, by modifying the mean field assumptions -- considering, for example,
a non-trivial network of connections $J_{ij}$ -- one could also obtain a
solution for a liberal economy, $\alpha \le 1$ \cite{bm}.

If one calculates the average of the distribution (\ref{pls}), which
corresponds to the average wealth of the individual, one sees that the basic
difference between a social and a liberal economy is that it is finite in
the former case and infinite in the latter. Thus, for $\alpha \le 1$ one
would, due to the non-integrable tail of the distribution, expect the
appearance of rich individuals in the ensemble, with a wealth $N^{1/\alpha}$
times larger than the typical value. The authors of \cite{bm} interpreted
this result as a condensation phenomenon.

{\bf 3.\,\,\,}
In reality, this is not the case and the total wealth
of the society $W$ is in general fixed, thereby upsetting overall
scale invariance and giving us 
a closed system of type (a). How would the condensation phenomenon change in this
case? One  way to address this issue is to solve the 
equations of the type (\ref{dyn}) on the 
hypersurface $W=w_1+...+w_N$. This problem is 
reminiscent of Kac's master equation~\cite{KAC} on the sphere 
(fixed energy) for which a factorizable and stationary solution was 
derived in the thermodynamic limit under mild assumptions. 

Here, we follow a more phenomenological treatment  
and assume that $p(w) \sim 1/w^{1+\alpha}$ characterizes
the single wealth-distributions in the ensemble (a) with the individual
wealths adding to $W = w_1 + ... + w_N$. In this way, we have an
asymptotic Pareto macro-economy with a factorizable N-distribution
of wealths constrained on the hypersurface  of fixed wealth $W$.
For convenience, we assume that each individual wealth $w_i$
is an integer given in units of the smallest available currency unit.
The joint probability distribution of $w_i$'s is:
\begin{equation}
P (w_1, . , w_N) \ = \ \frac{1}{Z (W, N)} \, \prod_i \, p(w_i) \,
\delta \left( W - \sum_{i = 1}^N w_i \right) \, ,
\end{equation}
where $Z (W, N)$ is the appropriate normalization factor,
\begin{equation}
Z (W, N) \ = \ \sum_{\{w_i \ge 0\}} \, \prod_i \, p (w_i) \,
\delta \left( W - \sum_{i = 1}^N w_i \right) \, .
\end{equation}
This model is known as the balls-in-boxes or backgammon model
\cite{bbj1} where it has been applied to various
condensation and glassy phenomena. It can be solved 
in the limit of an infinite number of boxes $N$ and fixed density
of balls per box $\rho=W/N$ (thermodynamical limit)
by introducing the integral representation of the
delta function
\begin{eqnarray}
Z(N,\rho) =&& \ \sum_{\{w_i \ge 0\}} \, \prod_i \, p (w_i) \
\nonumber\\ &&\times \frac 1{2\pi}\,\int\limits_{-\pi}^{\pi}\!\mbox{d}\lambda
e^{-i\lambda(w_1+\cdots+w_N-\rho N)}\nonumber\\
=&&\frac{1}{2\pi}\int\limits_{-\pi}^{\pi}\!\mbox{d}\lambda e^{i\lambda \rho N}
\left(\sum_{w}p(w)e^{-i\lambda w}\right)^N \nonumber\\
=&& \frac{1}{2\pi}\int\limits_{-\pi}^{\pi}\!\mbox{d}\lambda
\exp\left(N (i\lambda\rho+ K({i\lambda})\right)
\end{eqnarray}
where $K$ is a  generating function given by
$
K(\sigma)={\rm ln}\,\sum_{w=1}^\infty p(w)e^{-\sigma w}
$.
Evaluating the integral using steepest descent gives
\begin{eqnarray}
f(\rho) = \sigma_*(\rho)\rho+ K(\sigma_*(\rho))\,\,,
\end{eqnarray}
where $\sigma_*(\rho)$ is a solution of the saddle point equation
$\rho + K'(\sigma_*) = 0$
and $f(\rho)$ is a 
free energy density per box, 
$ Z(W,N)= e^{N f(\rho) + \dots}.$
For a suitable choice
of the weights $p(w) \sim 1/ w^{1+\alpha}$ the system displays
a two phase structure as
the density is varied with a critical density $\rho_{cr}$.
When $\rho$ approaches
$\rho_{cr}$ from below, $\sigma_*$ approaches $\sigma_{cr}$ from above.  When
$\rho$ is larger than $\rho_{cr}$~, $\sigma$ becomes equal
to the critical value $\sigma_{cr}$ and the free energy is linear in
$\rho$
\begin{eqnarray}	
f(\rho) = \sigma_{cr}\rho+ \kappa_{cr}\,\,,
\label{fcond}
\end{eqnarray}
where $\kappa_{cr}=K(\sigma_{cr})$. The change of regimes
at $\rho_{cr}$ corresponds to a condensation transition, in which
an extensive fraction of the balls is in a single box.
The critical value $\sigma_{cr}$ is equal to the logarithm
of the radius of convergence of the series in the generating
function $K(\sigma)$. In particular, for purely power-like
weights 
\begin{eqnarray}	
p(w) = \frac{1}{\zeta(1+\alpha)} w^{-1-\alpha} \ , \quad w=1,2,\dots, 
\label{plw}
\end{eqnarray}
$\sigma_{cr}=0$. The normalization factor is given
by the Riemann Zeta function.
At the end of this section we will
comment on the case when the radius of convergence
of the series $K(\sigma)$ differs from one.

\begin{figure}
\begin{center}
\psfrag{xx}{\small {$w$}}
\psfrag{yy}{\small {$\hat{p}(w)$}}
\includegraphics[width=8cm]{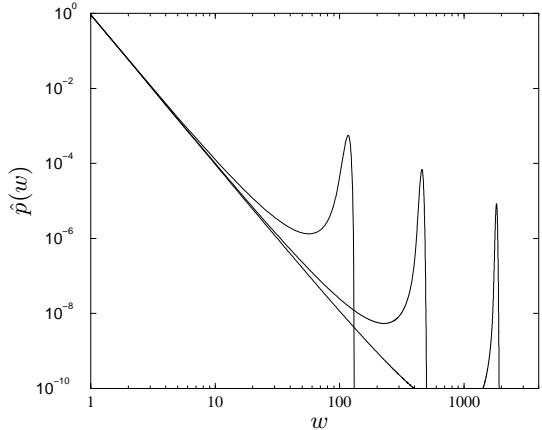}
\caption{\label{fig} Effective probability density of wealth. See text.}
\end{center}
\end{figure}

The transition to a condensed phase happens when ${W}/{N}$
becomes larger than a critical density $w_*$, which is nothing but the mean
wealth
\begin{equation}
w_* = \sum_{w} w \, p(w) \, .
\end{equation}
Since we can change the small $w$ part of the distribution by tuning the
appropriate macro-economical parameters without affecting the large $w$
behavior of $p(w)$, we have some control over where the
threshold $w_*$ will lie. We can define
an {\it effective} probability distribution of wealth:
\begin{equation}
\widehat{p}(w) = \frac{1}{N}\left\langle \sum_i^N \delta(w_i-w)
\right\rangle_P \, 
\end{equation}
which now, unlike the original $p(w)$, takes into account the finite total wealth $W$.
Below threshold $w_*$, the system is in a phase
in which the effective probability distribution $\widehat{p}(w)$
has an additional scale factor
in comparison with the old distribution $p(w)$
\begin{equation}
\widehat{p}(w) \sim e^{-\sigma w} p(w) \, .
\label{fluid}
\end{equation}
Here, $\sigma$ depends only on the difference ${W}/{N} - w_*$. It
vanishes at  threshold, so that the old Pareto tails are restored at
this point. Above  threshold, the macro-economy responds to the
increasing average wealth by creating a single individual with a wealth
proportional to the total wealth $W$, namely $w_{max} = W - N w_*$
\begin{equation}
\widehat{p}(w) \sim p(w) + \frac{1}{N} \delta_{w, W - N w_*} \, .
\end{equation}
The behavior of $\hat{p} (w)$ versus $w$ is shown in Fig.~\ref{fig},
for index $\alpha=3$, $N=128, 512, 2048$ and a density $W/N >w_*$.
At threshold the inverse participation ratio
\begin{equation}
Y_2 = \frac{1}{N^2} \left\langle \sum_i w_i^2 \right\rangle_P
= \frac{1}{N} \sum_w w^2 \widehat{p}(w)\,\,,
\end{equation}
changes, in the large N limit, from 0 to $({W}/{N} - w_*)^2$,
signaling the appearance of a wealth condensation. Basically, everything
in excess of the critical wealth $N w_*$ ends up in the portfolio of a
single individual. We call this the surplus anomaly. It can appear only
in a social economy ($\alpha > 1$), because only in this case do we have
a finite critical wealth per individual $w_*$.

In a liberal economy, $w_*$ is obviously infinite, meaning that the
system remains always below threshold and there is never any
condensation. Note that these results for the closed model (a)  do not contradict the results of the
previous section for the super-open model (c) since we now have a well-defined average
wealth ${W}/{N}$ which prevents the appearance of individuals with
a wealth $w \sim N^{1/\alpha}$ growing faster than linearly.

The behavior we have discussed here 
for closed systems
is not restricted to power-law
weights $p(w)$. The saddle point equation for 
the generating function $\rho + K'(\sigma_*) = 0$, 
can  have similar properties
for other functional forms of the weights. For instance,
one can easily check that a change of weights
$p(w) \rightarrow e^{-\bar{\sigma} w} p(w)$
merely leads to a change 
$\sigma_{cr}\rightarrow \sigma_{cr}+\bar{\sigma}$
leaving the phase structure of the model intact.
In particular, if the weights (\ref{plw}) had
an exponential pre-factor $p(w) \sim e^{-\bar{\sigma} w}/ w^{1+\alpha}$, 
we would have $\sigma_{cr}=\bar{\sigma}$, but the critical
density:
\begin{eqnarray}
\rho_{cr} = \frac{\zeta(\alpha)}{\zeta(\alpha+1)}
\end{eqnarray}
would be independent of $\bar{\sigma}$. Clearly,
the critical properties of the model
are encoded in the sub-exponential behavior of the
weights $p(w)$ for large $w$. 
Solving the saddle point equation one can
check that for weights with power-like sub-exponential
behavior, the most-singular part of the free energy
$f(\rho)$ has a branch point singularity when 
$\Delta \rho = \rho_{cr}-\rho \rightarrow 0^+$:
\begin{eqnarray}
f(\rho) = \left\{\begin{array}{lll} 
\Delta \rho^{\alpha/(\alpha-1)} & \mbox{for} & 1<\alpha<2 \\
\Delta \rho^{{\alpha}} & \mbox{for} & \alpha\ge 2 
\end{array}\right. \ .
\label{sing}
\end{eqnarray}
For integer values, the power-like singularity changes
to a singularity of the type integer power times logarithm.

One may consider other functional sub-exponential forms
of the weights $p(w)$. A criterion for the
presence of the phase transition
is that the derivative of the generating function
is finite, $-K'(\sigma_{cr}) < \infty$, at the radius of 
convergence $\sigma_{cr}$. Physically this means
that the critical density is finite.
For example, stretched exponential weights 
\begin{eqnarray}
p(w) \sim e^{- \beta w^\delta}, 
\label{se}
\end{eqnarray}
with $0<\delta <1$ and $\beta>0$ have this property.
As before, we have a saddle point phase for small density $W/N$,
with an exponential suppression of large wealths, and a
condensed phase for large density $W/N$, with a surplus anomaly. 
At the transition point, however, instead of the Pareto
distribution we have (\ref{se}). The second derivative
of the free energy is discontinuous at the transition.
If the transition is approached from the condensed phase
$\Delta \rho = \rho_{cr}-\rho \rightarrow 0^-$, 
$f''(\rho)=0$, while from the saddle point one
$\Delta \rho \rightarrow 0^+$:
\begin{eqnarray}
f''(\rho) = \sigma'_*(\rho) = -\frac{1}{K''(\sigma_{cr})} \, .
\end{eqnarray}
For the weights (\ref{se}) as well as the power-like 
weights for $\alpha>2$, the derivative $K''(\sigma_{cr})$ 
is finite. Thus in both  cases 
the second derivative of the free energy is 
discontinuous. In contrast, for $1<\alpha<2$,  
$K''(\sigma_{cr})=\infty$ and $f''(\rho)=0$ on
both  sides of the transition. In this case
the singularity yielding the discontinuity
of derivatives of the free energy is given by 
(\ref{sing}). The transition becomes arbitrarily soft
when $\alpha\rightarrow 1$.

{\bf 4.\,\,\,}
Finally, let us discuss an economy in contact with one or more external
ones (ensemble (b)). The total wealth $W$ is not fixed in this case, but
may adjust dynamically to an equilibrium value given by a stability
parameter $\mu$ (inverse temperature). The
partition function for this ensemble is given by \cite{bbj2}
\begin{equation}
{\cal Z} (\mu, N) = \sum_W Z (W, N) \, e^{-\mu W} \, .
\end{equation}
The total wealth in our economy now depends on the value of $\mu$.
The model has a phase transition at $\mu = 0$. For $\mu > 0$, the average
wealth per individual ${W}/{N}$ fluctuates according to a Gaussian
distribution with a certain average value $w_* (\mu)$ and a width that
is inversely proportional to the square root of the system size
${1}/{\sqrt{N}}$. At the critical point $\mu = 0$, the situation
becomes unstable as the economy starts to attract the attention of
the outside world and $W$ acquires a tendency to grow. In an idealized
situation where the outside world has limitless wealth, ${W}/{N}$
would actually become infinite as soon as $\mu < 0$. In practice, of
course, it remains bounded by an upper limit.

The order parameter for this transition is $r = {N}/{W}$, which in
the idealized case is zero for $\mu < 0$ and positive otherwise. Its
critical behavior depends on $\alpha$ as
\begin{equation}
r \sim \mu^{1/\alpha} \quad {\rm for} \ \mu \rightarrow 0^+ \, .
\end{equation}
The order of the transition thus depends on the type of our economy.
In a social economy ($\alpha > 1$), the transition is of first order
and $r$ changes discontinuously at the critical point. In a liberal
economy ($\alpha \le 1$), the transition is continuous, and becomes
arbitrarily soft as $\alpha$ approaches zero.

The $r = 0$ phase is one where condensation takes places not only within
the considered economy, but in the whole system including the outside world.
To better illustrate this situation, consider two mean-field Pareto
economies, each with the same distribution $p(w)$ but possibly different
total wealths $W_1$ and $W_2$. If we bring them into contact with each
other, they will form a larger mean-field economy with a constrained
total wealth $W = W_1 + W_2$. For $\mu = 0$, condensation can take place
with equal probability in either one of them, so if we look at only one
of the systems, we might observe condensation or we might not. In other
words, there are large fluctuations. But if $\mu \ne 0$, then one of the
subsystems will favor condensation, and wealth will tend to flow towards
it. The other system then has to adjust to the fact that wealth disappears
from it. This leads exactly to the two phases discussed above.


{\bf 5.\,\,\,}
We have shown that in
a social economy, condensation may occur if the total wealth of 
the society exceeds a certain critical value. In our analysis, the 
system favors the occurrence of a single individual in possession 
of a finite fraction of the economy's total available wealth,
providing a physical mechanism for ``corruption''. The analysis
we have provided may be improved by considering (\ref{dyn}) in
general, using a random network for the flow channels restricted
to a hypersurface of fixed wealth. In this way, we could learn more 
about the statistical aspects underlying the process of fortune 
creation and propagation. \\\\
{\bf Acknowledgments:}
We would like to thank Ewa Gudowska-Nowak for a critical discussion on 
stochastic analysis.
This work was partially supported by
the EC IHP network {HPRN-CT-1999-000161},
the Polish-British Join Research Collaboration
Programme under the project WAR/992/137,
the Polish Government Projects (KBN) 2P03B 00814, 
2P03B 01917, the Hungarian FKFP grant 220/2000,
and the US DOE grants DE-FG02-88ER40388.

\end{document}